\documentclass[twocolumn,showpacs,amsmath, amssymb]{revtex4}

\usepackage[unicode]{hyperref}
\usepackage[caption=false]{subfig}
\usepackage{graphics}
\usepackage{epsfig}
\usepackage{graphicx}
\graphicspath{ {./images/} }
\usepackage{dcolumn}
\usepackage{bm}
\usepackage[caption=false]{subfig}
\usepackage[utf8]{inputenc}
\usepackage{dirtytalk}
\usepackage[autostyle]{csquotes}
\usepackage{xcolor}

\begin{document}

\title{
A  weighted planar stochastic lattice with scale-free, small-world and multifractal properties
}

\author{Tushar Mitra and Md. Kamrul Hassan
}
\date{\today}

\affiliation{
Theoretical Physics Group, Department of Physics, University of Dhaka, Dhaka 1000, Bangladesh
}

\begin{abstract}%
We investigate a class of weighted planar stochastic lattice (WPSL1) created
by random sequential nucleation of seed from which a crack is grown parallel to
one of the sides of the chosen block and ceases to grow upon hitting another crack.
It results in the partitioning of the square into contiguous and non-overlapping 
blocks. Interestingly, we find that the dynamics of WPSL1 is governed by infinitely 
many conservation laws and each of the conserved quantities, except the trivial conservation
of total mass or area, is a multifractal measure. 
On the other hand, the dual of the lattice is a scale-free network as its degree distribution 
exhibits a power-law $P(k)\sim k^{-\gamma}$ with $\gamma=4.13$. The network is also a 
small-world network as we find that (i) the total clustering coefficient $C$ is high and 
independent of the network size and (ii) the mean geodesic path length grows logarithmically with $N$. Besides, the clustering coefficient $C_k$ of the nodes which have degree $k$ decreases exactly as $2/(k-1)$ revealing that it is also a nested hierarchical network. 
\end{abstract}



\maketitle

\section{Introduction}

Structures, especially lattice-like structures, are important in physics as we often use them as 
a backbone while solving theoretical models to gain deeper insight into various physical 
phenomena. These structures, however, must have some universal 
properties such as the number of nearest neighbors, next nearest neighbors, symmetry etc.
which are characteristic hallmarks of a given structure. The success of a theoretical model often
depends not only on how good the model itself is but also on the properties of the backbone
on which the model is applied.  For instance, we use regular and ordered structures like Bravais lattice, honeycomb lattice, triangular lattice etc. as a backbone to solve models that help us knowing 
the properties of matter since life-less atoms assemble themselves in a regular and ordered structure in 
the solid-state phase. However, there are also models to study the spreading of contagious diseases, 
rumors,  information, biological and computer virus that occur not through regular structure rather through random or scale-free and/or small-world network. Recently, topological network structures have been gaining popularity since many biologically motivated physics models on this structures are studied more than ever before
\cite{ref.ising_1,ref.ising_2,ref.sandpile}. In fact, we are now living in a world which is facing a pandemic due to COVID-19 and it has resulted
in a surge of research interest in epidemiology \cite{ref.epidemic_vespignani,ref.infection,ref.sir, ref.barabasi_decade,ref.Cattuto}. Thus, scale-free, small-world network as a backbone is more 
relevant now than ever.

Prior to 1998, topological network as a structure and its properties have been studied only by mathematicians. 
Two seminal works on scale-free and small-world network in the
late 1990s have revolutionized the notion of network \cite{ref.barabasi_science,ref.watts}. 
Soon scientists have found that most real-life networks through which disease, rumor, 
messages, viruses etc. spread are neither completely regular nor completely random but rather scale-free
in nature and often possess small-world properties too \cite{ref.vespignani,ref.newman,ref.meyers}. 
One of the characteristics of nodes in the topological network is their degree $k$ defined 
by the number of other nodes to whom a given node is connected. In the case where nodes in the 
network have great many different degree $k$, it is often worth investigating the degree distribution
 $P(k)$ which quantifies the probability that a node picked at random have degree $k$. 
The hallmark of the  Erd\"{o}s-R\'{e}nyi (ER) model is that $P(k)$ follows Poisson distribution 
suggesting that it is almost impossible to find nodes that have significantly higher or fewer links 
than the average degree \cite{ref.erdos}. On the other hand, scale-free network exhibits a power-law
$P(k)\sim k^{-\gamma}$ with exponent $\gamma$ whose value is mostly found approximately within $2$ and $4$ \cite{ref.www, ref.newman_book, ref.barabasi_review, ref.barabasi_physica}. 
The mean geodesic path length $l$, the average of the shortest paths among all the connected 
pairs of nodes, is yet another quantity that characterizes the long-range 
properties of the network \cite{ref.bollabas}. 
 Watts and Strogatz in 1998 introduced the idea of clustering coefficient which can be considered as yet
another characteristic of nodes in the network \cite{ref.watts}. The clustering coefficient
$C_i$ of a node $i$ is defined as the ratio of the number of edges $e_i$ that actually exist among 
its $k_i$ neighbors to the maximum possible edges $z_i=k_i(k_i-1)/2$ that could exist among the 
same $k_i$ neighbors. The clustering co-efficient $C$ of the whole
network is then obtained by averaging it over the entire network.

Earlier in 2010 we proposed a weighted planar stochastic lattice (WPSL), which we now regard as WPSL2,
and have shown that despite the coordination number disorder and the block size disorder 
it still exhibits some order \cite{ref.hassan_njp, ref.hassan_jpc}. Clearly, the construction process of 
WPSL2 would result in a fractal if one of the blocks with certain rules were
removed from the system like in the construction of Cantor set or in Sierpinsky carpet
\cite{ref.multifractal_1}. In one dimension, such systems are fractal and exhibit simple scaling but in higher
dimensions they are multifractal and exhibit multiscaling \cite{ref.krapivsky_pla,ref.fractal_1}. 
Recently, we have shown that the emergence of stochastic fractal is also accompanied by a conservation 
law which is reminiscent of Noether's theorem \cite{ref.noether_1, ref.noether_2}. This is also true for multifractal. In the
case of WPSL2, it is governed by infinitely many conserved quantities 
$\mu_i=x_i^{m-1}y_i^{4/m-1}$ $\forall \ m$
where $x_i$ and $y_i$ are the length and width of the $i$th block. For each value of $m$,
 except $m=2$ which corresponds to trivial conserved quantity namely conservation of total area, 
if the value of $\mu_i$ is distributed in the $i$th block then the total content $M=\sum_i\mu_i$
is unevenly distributed across the square (initiator). We have shown that such uneven distribution 
can be best quantified as multifractal. The support is a two dimensional Euclidian space where
the conserved quantities are distributed. Each of the non-trivial conserved quantities gives rise
to multifractal spectrum. 
The strongest criticism about that model was, however, that the exponent of the degree distribution $\gamma$  was far too high compared to the most empirical data or theoretical model has so far predicted. Besides, we do not yet know whether it possesses small-world properties or not. The impact of such scale-free and
small-world properties has already been proved to leave its signature through distinct results which 
were otherwise unexpected \cite{ref.wpsl_percolation_1,ref.wpsl_percolation_2,ref.dayeen_csf} . 

In this article, we propose a close variant of the weighted planar stochastic lattice. Like WPSL2, 
it too describes the partition of a square into  contiguous and 
mutually exclusive rectangular blocks. Here at each time step, a seed is nucleated randomly on a square  whereby a crack parallel to one of the sides of the square is grown 
and then ceases to grow upon hitting another crack. We name this as WPSL1 as it is created 
by one crack at each time step and WPSL2 when it is created by two orthogonal cracks at each time step.
We show that WPSL1 is governed
by infinitely many conservation laws and each non-trivial conserved quantity is a multifractal measure.
The dual of the WPSL1,
obtained by replacing the center of each block with a node and the common border between two blocks
by an edge connecting the corresponding nodes, can be described as a network. We can regard  WPSL1
as growing by sequential addition of one node and WPSL2 as a network growing by sequential addition of a group of three nodes \cite{ref.hassan_njp}. We show that the dual of the WPSL1 too self-organizes into a scale-free network but with exponent much smaller than that of WPSL2. 
Finally, we study the clustering coefficient and mean geodesic path length and find that the dual of both
WPSL1 and WPSL2 are  nested hierarchical network and WPSL1 is also truly small world. 

The organization of the rest of this paper is as follows. In section II, we revisit the topological and geometric properties of weighted planner stochastic lattice with two cracks (WPSL2) to coordinate with the newly proposed lattice. In section III, we propose a stochastic planar lattice with one crack which we call WPSL1 and describe the algorithm. In section IV, various topological properties of WPSL1 are discussed and we show that its dual is a scale-free network with exponent of the degree distribution much lower than 
that of the WPSL2. In section V, the various geometric properties of WPSL1 is explored and we show that it is a multi-multifractal like WPSL2. In section VI, the small world properties of WPSL1 and WPSL2 are discussed and compared. Finally, results are discussed and conclusions drawn in section VII.

\section{Weighted planar stochastic lattice by two cracks: WPSL2}

In 2010 we proposed a weighted planar stochastic lattice  that provided many non-trivial topological and geometric properties \cite{ref.hassan_njp}. We considered that the substrate is a square of unit 
area and at each time step a seed is nucleated from which two orthogonal partitioning lines, 
parallel to the sides of the substrate, are 
grown until intercepted by existing lines. It results in partitioning the square into ever smaller mutually 
exclusive rectangular blocks. The condition is that, the higher the area of a block, the higher is the probability that
the seed will be nucleated in it and divide it into four smaller blocks since seeds are sown at random on the substrate.
The resulting lattice, which we now call weighted planar stochastic lattice or in short WPSL2, 
where the factor $2$ refers to the fact that two mutually perpendicular partitioning lines 
are used to divide blocks randomly into four smaller blocks, has much richer properties than the square
and kinetic square 
lattice \cite{ref.hassan_njp,ref.hassan_jpc}. It provides an awe-inspiring perspective of 
intriguing and rich pattern of blocks which have different sizes and have great many different number of neighbors 
with whom they share common border. Such seemingly disordered lattice will have no place in physics 
unless there is some order and emergent behaviors from the statistical perspective. 
One of the emergent behavior is that the coordination number distribution function of WPSL2 follows inverse power-law \cite{ref.hassan_njp}. 
It immediately implies that the degree distribution of the network 
corresponding to its dual follows the same inverse power-law $P(k)\sim k^{-\gamma}$ with the same exponent $\gamma=5.66$.  
We have also shown that it is governed by infinitely many conservation laws and one of conserved
quantity can be used as multifractal measure so that the size disorder of the lattice can be quantified by multifractal 
scaling. Later Dayeen and Hassan have shown that each of the infinitely many conservation laws is actually a multifractal measure and hence WPSL2 is a multi-multifractal \cite{ref.dayeen_csf}.

The question is, can the two fundamental mechanisms, the growth and the preferential attachment rule, 
also be found responsible for the self-organization of the WPSL2
into a power-law degree distribution? Indeed, it was found that both the ingredients are 
present in WPSL2. However, while the presence of growth mechanism is inherent to the definition of the model, the 
preferential attachment rule is not as straightforward as that. For instance, at each time step a triad (three nodes linked by 
two edges) joins the existing network  and hence it is not static rather grows by sequential addition of a triad. 
It is interesting to note that it is not the node or the block which is picked preferentially 
with respect to area to gain links rather its neighbors. 
Yet, it embodies the preferential attachment rule since the higher the number of links
(neighbors) a given node (block) has, the higher is the chance that one of its neighbors will be picked and 
as a result will gain links with incoming nodes (blocks).
However, the small world properties, that are the clustering coefficient and the geodesic path length of WPSL2 have not been explored in any previous endeavors. 

\section{Weighted planar stochastic lattice by a single crack: WPSL1}

What if we further modify the generator of WPSL2 so that it now divides the initiator 
into two blocks either 
horizontally or vertically with equal {\it a priori} probability? It describes
a process whereby a seed is nucleated randomly at each time step on the initiator and 
upon nucleation a single line grows either horizontally or vertically with 
equal probability until intercepted by already existing lines.
Perhaps an exact algorithm rather than a mere definition can better describe the model.
In step one, the generator divides the initiator,
say a square of unit area, either horizontally or vertically into two smaller blocks at random. 

\begin{figure}
\centering

\subfloat[]
{
\includegraphics[height=7.0 cm, width=8.0 cm, clip=true]
{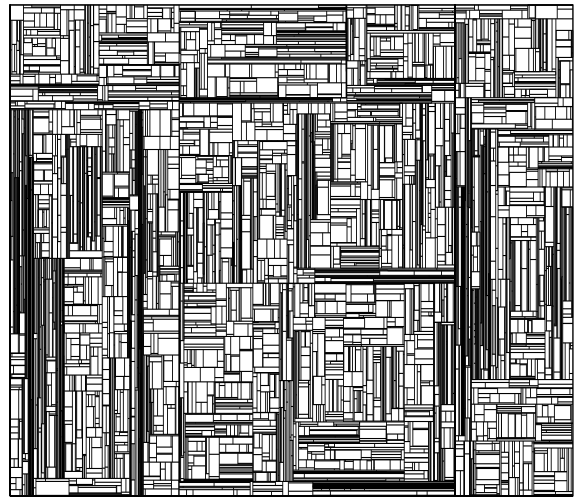}
\label{fig:1a}
}

\subfloat[]
{
\includegraphics[height=7.0 cm, width=8.0 cm, clip=true]
{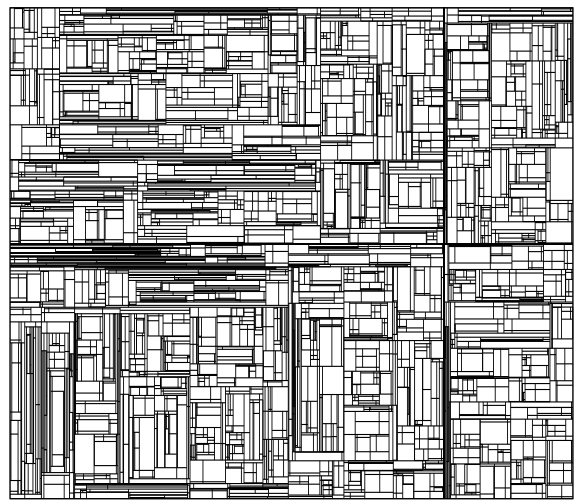}
\label{fig:1b}
}
\caption{ Snapshots of (a) WPSL1 (b) WPSL2 where both snapshots contain 4999 cells.} 

\label{fig:1ab}
\end{figure}

We then label the top or left block by $a_1$ if divided horizontally or vertically respectively and the other block
as $a_2$. In each step thereafter
only one block is picked preferentially with respect to their area (which we also refer to as the 
fitness parameter) and then it is divided randomly into two blocks in the same fashion i.e. either horizontally or
vertically. In general, the $j$th step of the algorithm can be described as follows.
\begin{itemize}
\item[(i)] Subdivide the interval $[0,1]$ into $j$ subintervals of size $[0,a_1]$, $[a_1, a_1+a_2]$,$\  ...$, 
$[\sum_{i=1}^{j-1} a_i,1]$ each of which represents the blocks labeled by their areas $a_1,a_2,...,a_{j}$ respectively.
\item[(ii)] Generate a random number $R$ from the interval $[0,1]$ and find which of the $j$ sub-intervals 
contains this $R$. The corresponding block it represents, say the $k$th block of area $a_k$, is picked.  
\item[(iii)] Calculate the length $x_k$ and the width $y_k$ of this block and keep note of the coordinate of the
lower-left corner of the $k$th block, say it is $(x_{low}, y_{low})$.
\item[(iv)] Generate two random numbers $x_R$ and $y_R$ from $[0,x_k]$ and $[0,y_k]$ respectively and hence
the point $(x_{R}+x_{low},y_{R}+y_{low})$ mimics a random point chosen in the block $k$.
\item[(v)] Generate a random number $p$ within $[0,1]$. 
\item[(vi)] If $p<0.5$ then draw a vertical line else a horizontal line through the point $(x_{R}+x_{low},y_{R}+y_{low})$ 
 to divide it into two smaller rectangular blocks. The label $a_k$ is now redundant and hence
it can be reused.
\item[(vii)] Label the left or top of the two newly created blocks as $a_k$ depending
whether a vertical or horizontal line is drawn respectively and the remaining block is then labeled as $a_{j+1}$.
\item[(viii)] Increase time by one unit and repeat the steps (i) - (vii) {\it ad infinitum}.
\end{itemize}
 
We name the resulting weighted planar stochastic lattice as WPSL1. A snapshot of this lattice too provides an awe-inspiring perspective of 
intriguing and rich pattern of blocks and its blocks also have different sizes 
and have great many different number of neighbors with whom they share common border, see Fig. (\ref{fig:1ab}). The number
of blocks $N(t)$ in WPSL1 grows linearly as $N(t)=1+\alpha t$ with $\alpha=1$ while in the
case of WPSL2 it grows in the same fashion except we have $\alpha=3$. 
It implies that the dual of the WPSL1 grows by sequential addition of a monad or a single node while 
the dual of the WPSL2 grows by sequential addition of triad or a group of three nodes already linked
by two edges.

\section{Topological properties of WPSL1}

We will first investigate the property of coordination numbers of the blocks with time. The number of blocks and the coordination numbers evolve with each step. If we define each step of the construction process as one unit of time, then we find a relation between the coordination number of the blocks with time. Note that, while calculating the coordination number, we are taking periodic boundary condition in account. The total number of blocks $N(t)$ grows with time linearly as $N(t) = 1+ \alpha t$ with $\alpha=1$. Now, the number of blocks having coordination number $k$ is given by $N_{k}(t)$ which also varies linearly with time as the plot of $N_{k}(t)$ vs $t$ results in a straight line passing through the origin as shown in Fig.(\ref{fig:2}). So, we can write $N_{k} = m_{k} t$. The ratio between $N_{k}(t)$ and $N(t)$ is then given by $\rho _{k}(t)$ = $N_{k}(t)/N(t)$ = $m_{k}$ if we consider $N(t)\sim$ $t$ in the long time limit. So the coordination number distribution $\rho_{k}(t)$ is independent of time for a fixed $k$ and can be evaluated by determining the slope of the resulting straight line of the plot of  $N_{k}$(t) vs $t$.
\begin{figure}
\centering

\includegraphics[width=8.5cm,height=8.0cm,clip=true]
{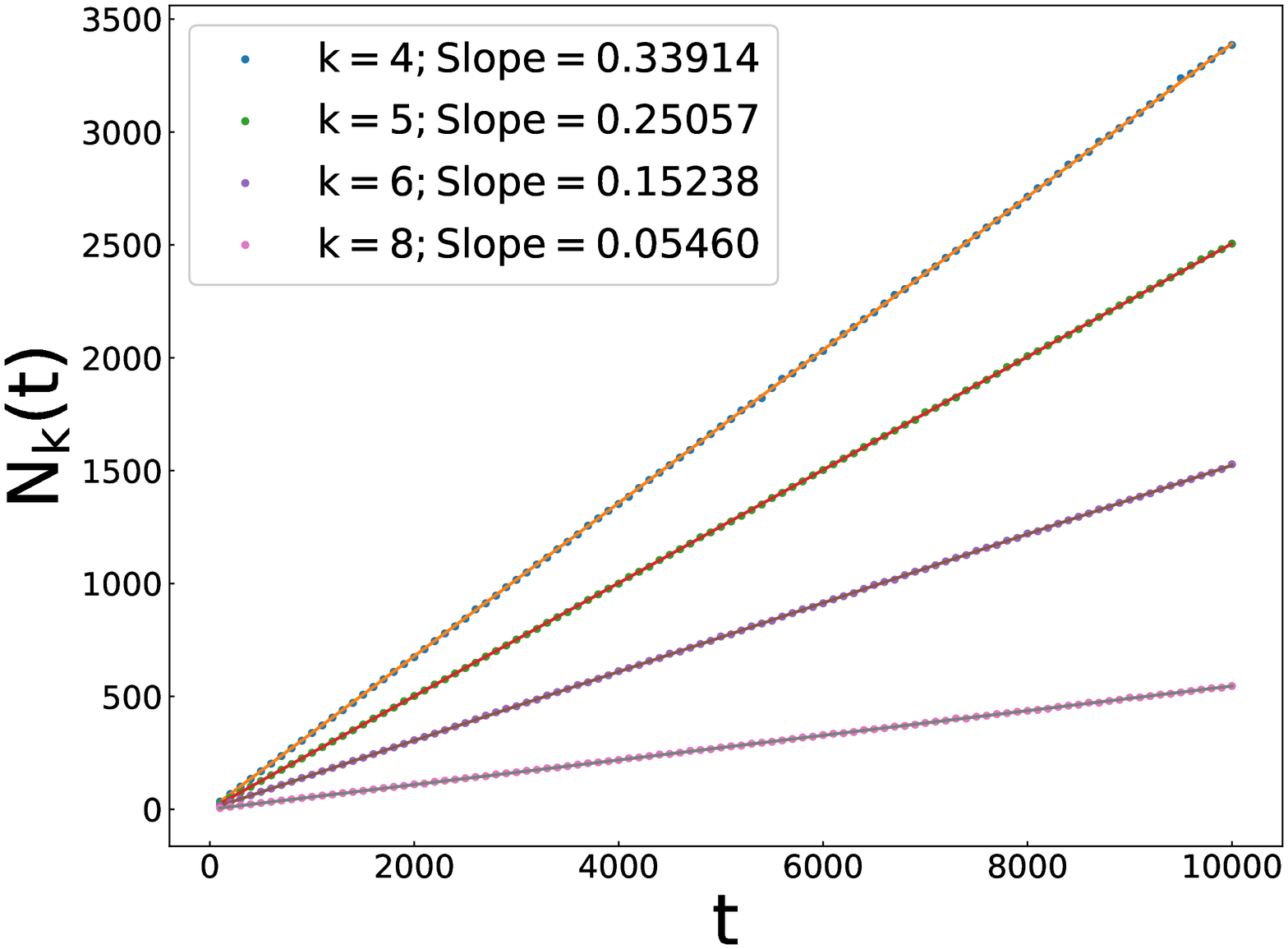}
\label{fig:2}

\caption{ Plot of the number of blocks $N_{k}(t)$ with k = 4, 5, 6 and 8 as a function of $t$. The slope of each straight line gives the coordination number distribution for the specific value of coordination number k = 4, 5 , 6 and 8.
}

\label{fig:2}
\end{figure}
However, for a fixed time, $\rho_{k}(t)$ becomes a function of coordination number $k$ and we simply denote it by $\rho(k)$. We can find the coordination number distribution $\rho(k)$ for a fixed time by finding the fraction of blocks having a coordination number exactly equal to $k$. Now, we can consider the blocks of the lattice as a node at the center of each block and the nearest neighboring blocks as other nodes connected by links. So, we can map the WPSL1 as a network, which is the dual of WPSL1 and we call it DWPSL1. It can be easily concluded that the degree distribution P(k) of this network will be equivalent to the coordination number distribution $\rho(k)$ of WPSL1. In Fig. (\ref{fig:3}), we plot the natural logarithm of the degree distribution ln($P(k)$) vs ln($k$). The resulting plot is a straight line with a slope equals to $-4.13$. This implies that the degree distribution obeys a power law given by,
\begin{equation}
\label{eq:degree}
P(k) \sim k^{-\gamma} ,
\end{equation}
with $\gamma=4.13$. This is much less than that of WPSL2 which was $5.66$. Also, the exponent found for WPSL1 is much closer to the real life networks than WPSL2. An interesting fact is that, this exponent is almost equal to the exponent of the degree distribution of the electric power grid, which is $\gamma=4$ \cite{ref.barabasi_science}. It is worth mentioning that the nodes of the electric power grid is spatially embedded just like the nodes of WPSL1.

\begin{figure}
\centering

\includegraphics[width=8.5cm,height=8.0cm,clip=true]
{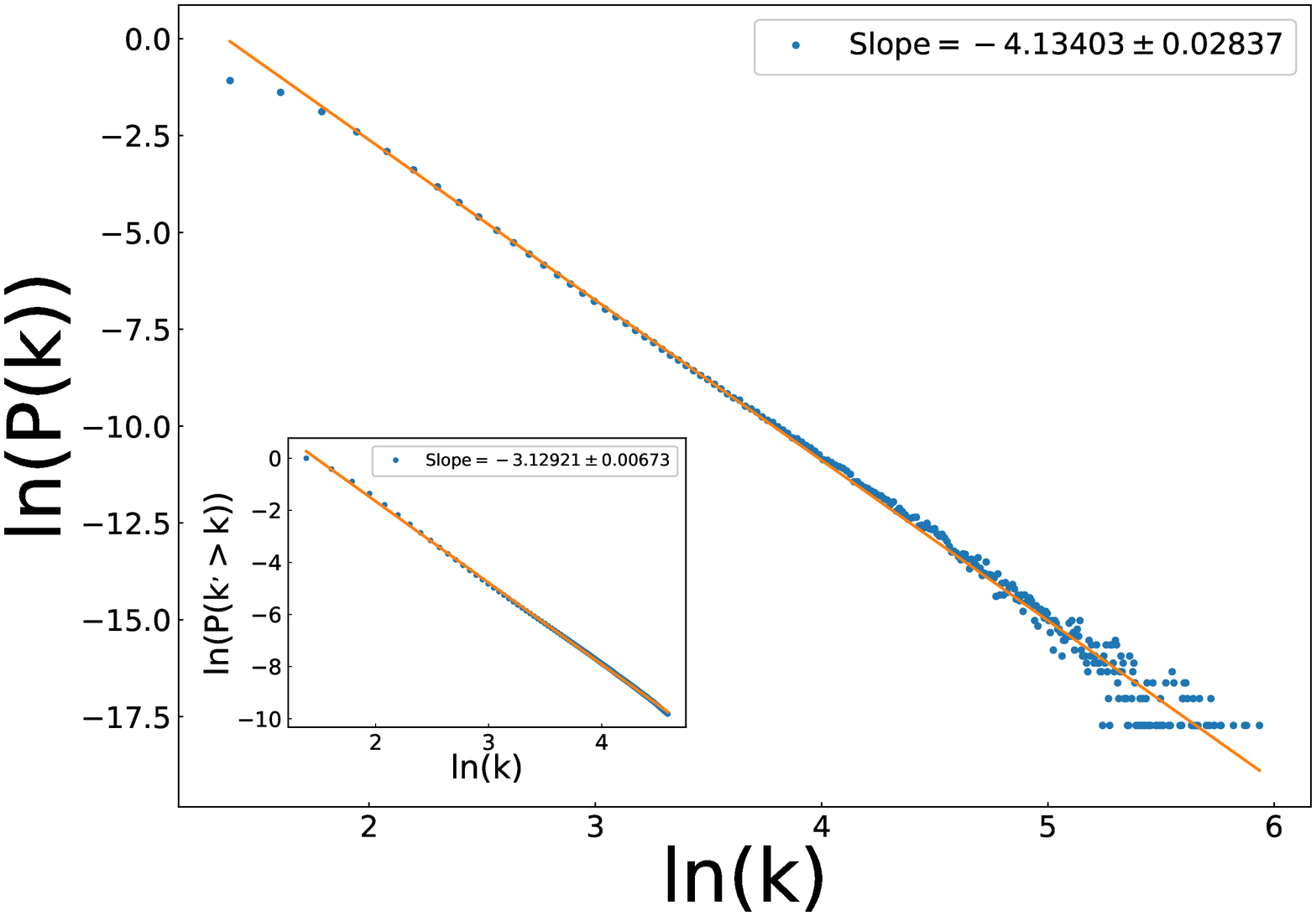}

\label{fig:3}
\caption{ Plot of $\ln(P(k))$ vs $\ln(k)$ of the DWPSL1 network for time $t = 10000$ where the data is averaged upon $5000$ independent realizations. The resulting plot is a straight line with a slope 
$\gamma= 4.13$ which reveals that $P(k)\sim k^{-\gamma}$. The cumulative degree distribution plot $\ln(P(k^{\prime}>k))$ vs $\ln(k)$ is shown in the inset which is again a straight line with a slope equals to $\gamma-1$ = $3.13$. 
}

\label{fig:3}
\end{figure}

One of the interesting features of the degree distribution plot is that it has a fat-tail, which is a significant characteristic of a scale free network. This means that there are few nodes with a very high number of connected neighbors, which are called $hubs$. These hubs play an important role for the spreading of physical quantities through a network. However, a heavy or fat tail in the degree distribution 
makes finding the exponent $\gamma$ a more problematic. To solve it we 
must invoke the idea of cumulative distribution $P(k^{\prime}>k)$ and plot $ln(P(k^{\prime}>k))$ vs $ln(k)$. The cumulative degree distribution is related to the degree distribution by the following law,
\begin{equation}
\label{eq:cumulative_degree}
P(k) = -\dfrac {dP(k^{\prime}>k)} {dk} .
\end{equation}
The plot of the cumulative distribution is expected to smooth out the fat-tail and result in a straight line in the infinite size limit. But, for the finite size of the network, the plot is a straight line up to a limit and then falls off. However, a straight line can be approximated in the finite size limit which has a slope greater than the slope of normal degree distribution plot by an additive factor of $1$. Indeed, by plotting the cumulative degree distribution (see the inset of Fig. (\ref{fig:3})) we can approximate a straight line which has a slope $-3.13$. So, it is evident from the plot that the degree distribution decays following a power law $P(k) \sim k^{-\gamma}$ with $\gamma = 3.13 + 1 = 4.13$.

The power-law degree distribution $P(k)$ has been found in many seemingly unrelated real life and man-made networks. It suggests that there must exist some common underlying mechanisms for which disparate systems behave in such a remarkably similar fashion \cite{ref.barabasi_review}. Barab\'{a}si and Albert in 1999 addressed exactly this and found that the growth and the preferential attachment rule are the 
main factors behind the emergence of such power-law degree distribution \cite{ref.barabasi_science}. Indeed, the dual of the WPSL network too grows with time. The dual of the WPSL2 grows by addition 
of a group of three nodes which are already linked by two edges. On the other hand, the dual 
of WPSL1, grows by addition of one new node. However, in either case the node which is picked at
random is not the one which gains links rather its neighbors gain link. It means the higher the number
of nearest neighbors, the higher the probability the block has to gain new neighbors {\it vis-a-vis} 
link. Using this idea, in 2017 we proposed a mediation-driven attachment (MDA) rule to construct Barab\'{a}si-Albert like network \cite{ref.hassan_liana}. At a glance, it may seem that MDA defies the PA
rule but a deeper look suggests that it embodies the intuitive idea of the PA rule, albeit in disguise. 
The MDA rule is in fact not only preferential but it also can be super-preferential in some cases.

\section{Geometric properties of WPSL1}

Now, we look into the geometric properties of WPSL1 in an attempt to check if there exists some rules and order. We can also use this model to describe the kinetics of planar fragmentation in an attempt to understand 
the extent of influence of size and shape when both are considered dynamical variables \cite{ref.hassan_pre,ref.krapivsky, ref.redner,ref.naim}. 
The distribution function
$f(x,y,t)$, that describes concentration of blocks of sides having length $x$ and width $y$ of WPSL1,
evolves as
\begin{eqnarray}
\label{eq:WPSL}
{{\partial f(x,y,t)}\over{\partial t}} & = & -xy f(x,y,t)+ {{1}\over{2}}\times 2 y\int_x^\infty f(x_1,y,t)dx_1  \nonumber \\ & + 
& {{1}\over{2}}\times 2  x\int_y^\infty f(x,y_1,t)dy_1.
\end{eqnarray}
The first term on the right accounts for the loss of blocks of sides $x$ and $y$ due to nucleation of seed. The pre-factor $xy$ here implies that the seeds are nucleated on the block chosen
preferentially according to the area of the existing blocks.
The second term  on the right describes placing a cut vertically   
with probability $1/2$ on a side $x_1>x$  of block of sides $x_1$  and $y$ 
to form two blocks, hence the factor $2$, of sides ($x, y$) and ($x_1-x,y$). Similarly, the third term represents the gain 
of blocks of sides $x$ an $y$ upon placing a cut 
horizontally with probability $1/2$ on a block of sides $x$ and $y_1>y$ to form two blocks of sides ($x, y$) and ($x,y_1-y$).

\begin{figure}
\centering

\includegraphics[width=8.5cm,height=8.0cm,clip=true]
{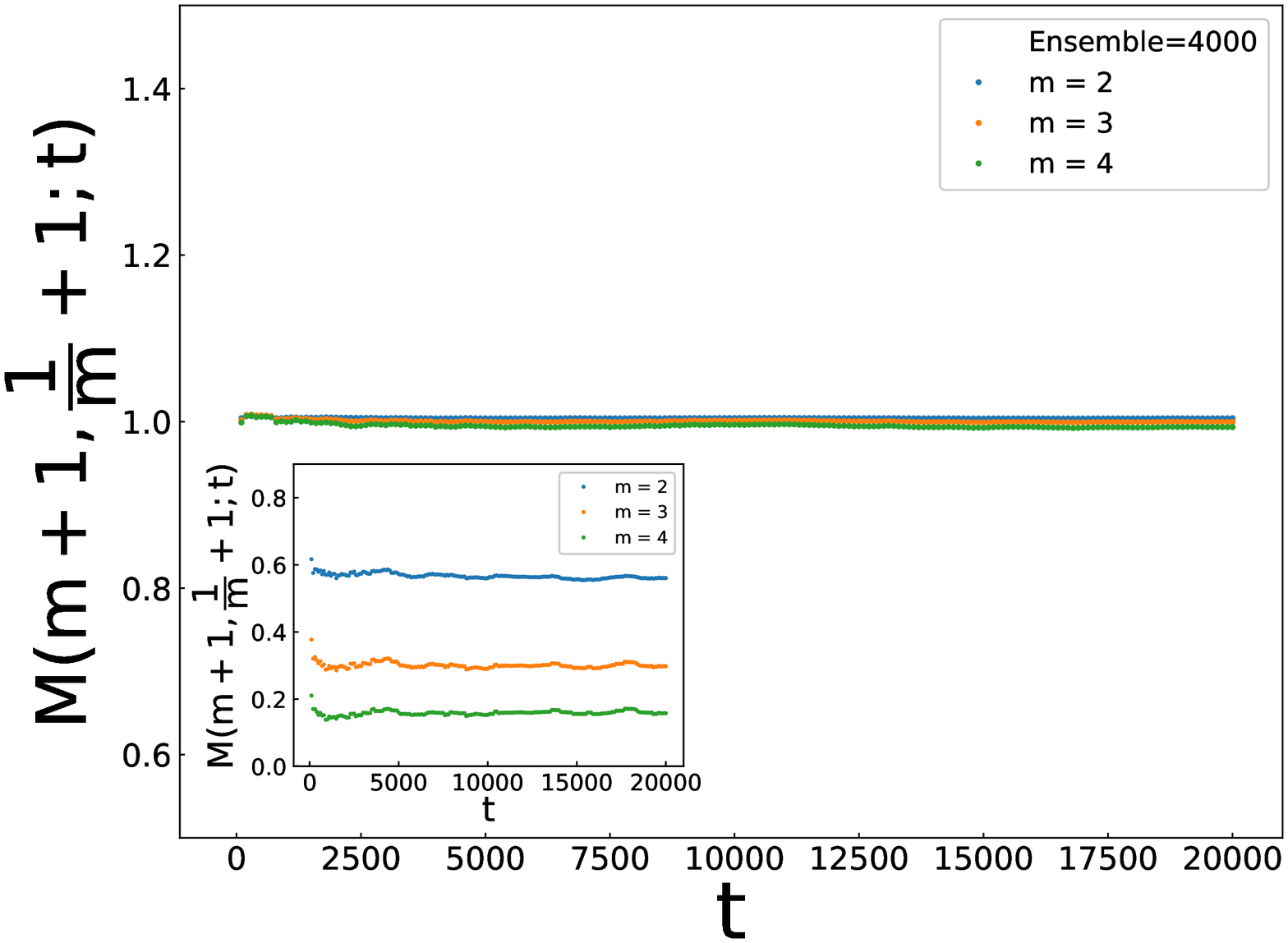}
\label{fig:4}

\caption{ Plots of the moment $M(m+1,1/m+1;t)$ vs $t$ for $m = 2, 3$ and $4$ for an average of $4000$ independent realizations. It is evident from the plot that the moments are constant with time and equal to $1$ for all values of $m$. The same plot for a single realization is shown in the inset
which shows that although $M(m+1,1/m+1;t)$ remains constant with time $t$,
the numerical value of the constants are different for each value of $m$.
}
\label{fig:4}
\end{figure}

 \begin{figure}
\centering

\includegraphics[width=8.5cm,height=8.0cm,clip=true]
{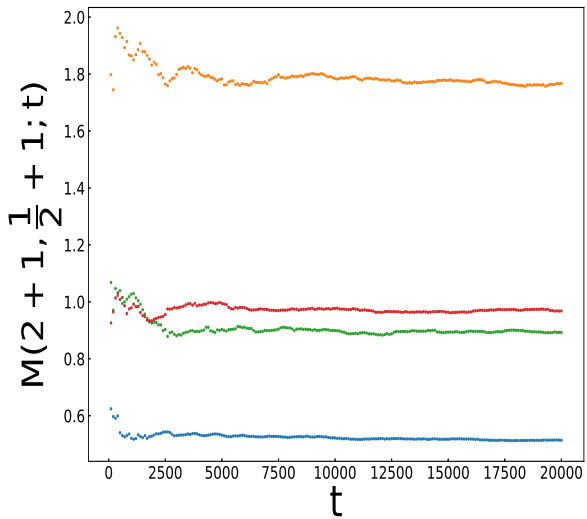}
\label{fig:5}

\caption{ Plots of the moment $M(m+1,1/m+1;t)$ vs $t$ for $m = 2$ for four different realizations. 
Although the values of $M(2+1,1/2+1;t)$ are constant with time for all the four realizations, 
the value of the moment is different for each realization.
}

\label{fig:5}
\end{figure}

We find it a formidable task to solve Eq. (\ref{eq:WPSL}) to find solution for $f(x,y,t)$. Instead,
we introduce the $2$-tuple Mellin transform of $f(x,y,t)$ given by
\begin{equation}
\label{momenteq_1}
{M(m,n;t)  =  \int_0^\infty \int_0^\infty x^{m-1} y^{n-1} f(x,y,t) dx dy,
}
\end{equation}
which describes moment at fixed $m$ and $n$. Incorporating it in Eq. (\ref{eq:WPSL}) yields
\begin{equation}
\label{momenteq_2}
{{dM(m,n;t)}\over{dt}} = \Big ( {{1}\over{m}}+{{1}\over{n}}-1\Big )M(m+1,n+1;t).
\end{equation}
It can be re-written as,
\begin{equation}
\label{momenteq_3}
{
{{dM(m,n;t)}\over{dt}} = - \Big ( {\alpha_{+}\alpha_{-}\over{m n}}\Big )M(m+1,n+1;t),
}
\end{equation}
where
\begin{equation}
\label{momenteq_4}
{
\alpha_{\pm}={{m+n-2}\over{2}}\pm\sqrt{ {{(m-n)^2}\over{4}}+1}.
}
\end{equation}
We can now iterate Eq. (\ref{momenteq_3}) to find various derivatives of $M(m,n;t)$ which are given by,
\begin{eqnarray}
\label{momenteq_5}
\dfrac {d^{j}M(m,n;t)} {dt^{j}} = (-1)^{j} \prod_{k = 1}^{j} \dfrac {(\alpha_{+}+k-1)(\alpha_{-}+k-1)} {(m+k-1) (n+k-1)} \nonumber \\ M(m+k,n+k;t) \nonumber \\
\end{eqnarray}
These derivatives  can now be used in Taylor series expansion of M(m,n;t) about $t = 0$ along with using mono-disperse initial condition which gives  \
\begin{equation}
\label{momenteq_6}
{
M(m,n;t) =  {}_{2}F_{2} (\alpha_{+},\alpha_{-};m,n;-t),
}
\end{equation}
where ${}_{2}F_{2}$ is the generalized hypergeometric function \cite{ref.hypergeometric}.

The nontrivial nature of the solution given by Eq. (\ref{momenteq_6}) can be best appreciated 
by considering its behavior in the limit $t\rightarrow \infty$ \cite{ref.hassan_multifractality_1}.
To this end, we find $M(m,n;t)$ decays following a power-law
\begin{equation}
\label{momenteq_7}
{
M(m,n;t) =  \dfrac {\Gamma(m) \Gamma(n) \Gamma(\alpha_{+}\alpha_{-})} {\Gamma(\alpha_{+})\Gamma(m-\alpha_{-})\Gamma(n-\alpha_{-})} t^{-\alpha_{-}}.
}
\end{equation}
It immediately implies that 
\begin{equation}
\label{momenteq_const}
M(m+1, 1/m+1;t)=const. \hspace{0.25 cm} \ \forall \ t >0,
\end{equation}
or if the blocks are labeled as $i=1,2,..., N$ then the WPSL1 obeys infinitely many conservation laws i.e.,
\begin{equation}
\label{consteq}
\sum_i^N x_i^my_i^{1/m}=const.,
\end{equation}
which includes the trivial conservation law that is conservation of total mass or area $M(2,2;t)=1$.

We performed extensive numerical simulation to measure $M(m+1,1/m+1;t)$ using Eq.(\ref{consteq}). We show
the plot of this in the inset of Fig. (\ref{fig:4}) as a function of $t$ for $m = 2, 3$ and $4$ 
for a single realization. The moment for each $m$ value remains constant with time albeit the numerical
values are different for every different values of $m$. Moreover, for a given value of $m$ the numerical
value of the conserved quantity is different for every single realization 
although it remains constant with time. The moment $M(m+1,1/m+1;t)$ for $m=2$ is plotted as a function of $t$ in Fig. (\ref{fig:5}) for four independent realizations, which confirms the above-mentioned statement. Interestingly, the ensemble average value of $M(m+1,1/m+1;t)$ is always equal to $1$ which is shown 
in Fig. (\ref{fig:4}) for $m=2, 3$ and $4$ where data is averaged over $4000$ independent realizations. This is not a surprise, since with some algebra the coefficient of $t^{-\alpha_{-}}$ in Eq.(\ref{momenteq_7}) turns out to be $1$ i.e., $M(m+1,1/m+1;t) = 1$  $ \forall \ m>0$.

We shall now invoke the idea of multifractality and show that each of the non-trivial 
conserved quantities (except the trivial
conserved quantity namely the conservation of total area) is distributed in the lattice in such
a fashion that each of them can be best described as a multifractal. 
We assume that $p_i=x_i^my_i^{1/m}$ is the probability that the $i$th block contains the total measure
$\sum_i^N x_i^my_i^{1/m}$ \cite{ref.hassan_multifractality_1,ref.hassan_multifractality_2}. We can now  
construct the partition function 
\begin{equation}
\label{partition_function}
Z_q=\sum_ip_i^q    
\end{equation} of the probability $p_i$ 
 and comparing it with the definition of the two-tuple Mellin transform  
of $f(x,y,t)$ we find that 
\begin{equation}
\label{partitioneq}
Z_q=M(mq+1, q/m+1;t).
\end{equation}
According to Eq. (\ref{momenteq_7}), the asymptotic solution for the partition function therefore is 
\begin{equation}
\label{partitioneq_2}
Z_q(t)\sim t^{\dfrac{\sqrt{(m-1/m)^2q^2+4}-(m+1/m)q}{2}}.
\end{equation} 
Measuring it using the square root of the mean area
\begin{equation}
\delta(t)=\sqrt{{{M(2,2;t)}\over{M(1,1;t)}}} \sim t^{-1/2},    
\end{equation} 
as an yard-stick  
we find  $Z_q$ decays following a power-law
\begin{equation}
\label{weightednumber}
Z_q(\delta)\sim \delta^{-\tau(q)},
\end{equation}
where the mass exponent 
\begin{equation}
\label{massexponent}
\tau(q)=\sqrt{(m-1/m)^2q^2+4}-(m+1/m)q.
\end{equation} 
The mass exponent $\tau(q)$ 
must possess the properties such that $\tau(0)$ is the Hausdorff-Besicovitch dimension of the support and $\tau(1)=0$ required by the normalization of the probabilities $p_i$, which in our case is clearly true since $\tau(0)=2$ and $\tau(1)=0$ \cite{ref.hassan_santo, ref.feder}.

\begin{figure}
\centering
\includegraphics[width=8.5cm,height=8.0cm,clip=true]
{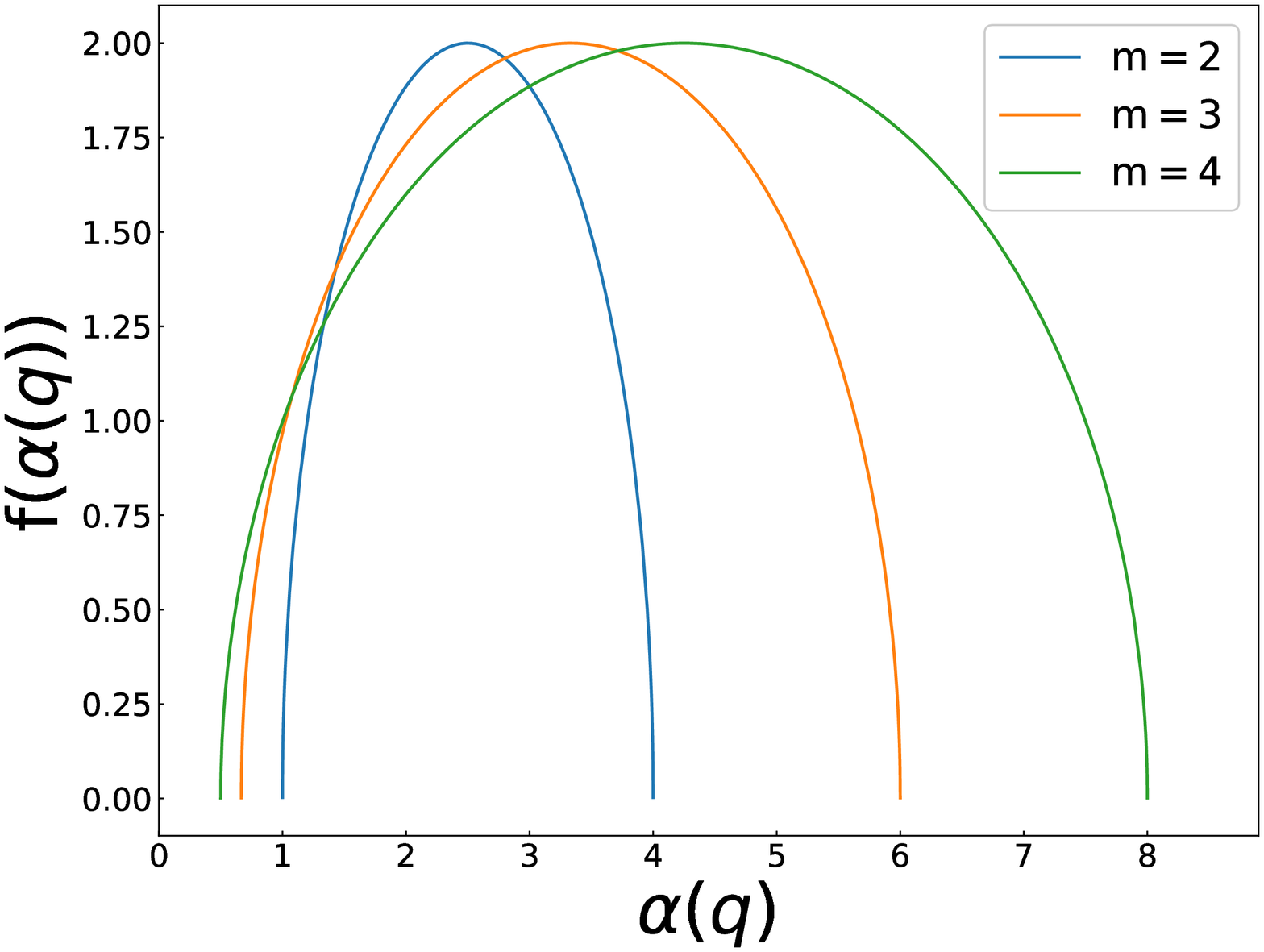}
\label{fig:6}

\caption{ Plot of the $f(\alpha)$ spectrum for $m = 2, 3$ and $4$. The maximum of each plot occurs
at $q = 0$ and the value is always equal to $2$. It indicates that the Hausdorff-Besicovitch dimension of the support is $2$. 
}

\label{fig:6}
\end{figure}

On the other hand, the Legendre transformation of the mass exponent $\tau(q)$ by using the Lipschitz-H\"{o}lder exponent $\alpha(q)$ is given by,
\begin{equation}
\label{Legendre}
f(\alpha)=\tau(q) + \alpha q,
\end{equation} 
with
\begin{equation}
\label{alpha_exponent}
\dfrac {d\tau} {dq} = -\alpha(q),
\end{equation} 
which gives,
\begin{equation}
\label{alpha_value}
\alpha(q)=(m+1/m)-\dfrac{q(m-1/m)^{2}}{\sqrt{(m-1/m)^{2}q^{2}+4}} .
\end{equation} 
$\alpha(q)$ now can be used as an independent variable which gives the multifractal spectrum 
\begin{equation}
f(\alpha(q))={{4}\over{\sqrt{(m-{{1}\over{m}})^2q^2+4}}}.
\end{equation}
It implies that except for the trivial mass conservation
law which corresponds to $m=1$ case, each and every other
conservation laws obtained by tuning the $m$ value results
in a spectrum of spatially intertwined fractal dimensions revealing the fact that WPSL1 exhibits multiple multi-fractality. Note that $f(\alpha(q))$ is always concave in character (see figure 6) with a single maximum $2$ at $q=0$ which corresponds to the dimension of the support.

\section{Small world properties of WPSL1 and WPSL2}

\begin{figure}
\centering

\includegraphics[width=8.5cm,height=8.0cm,clip=true]
{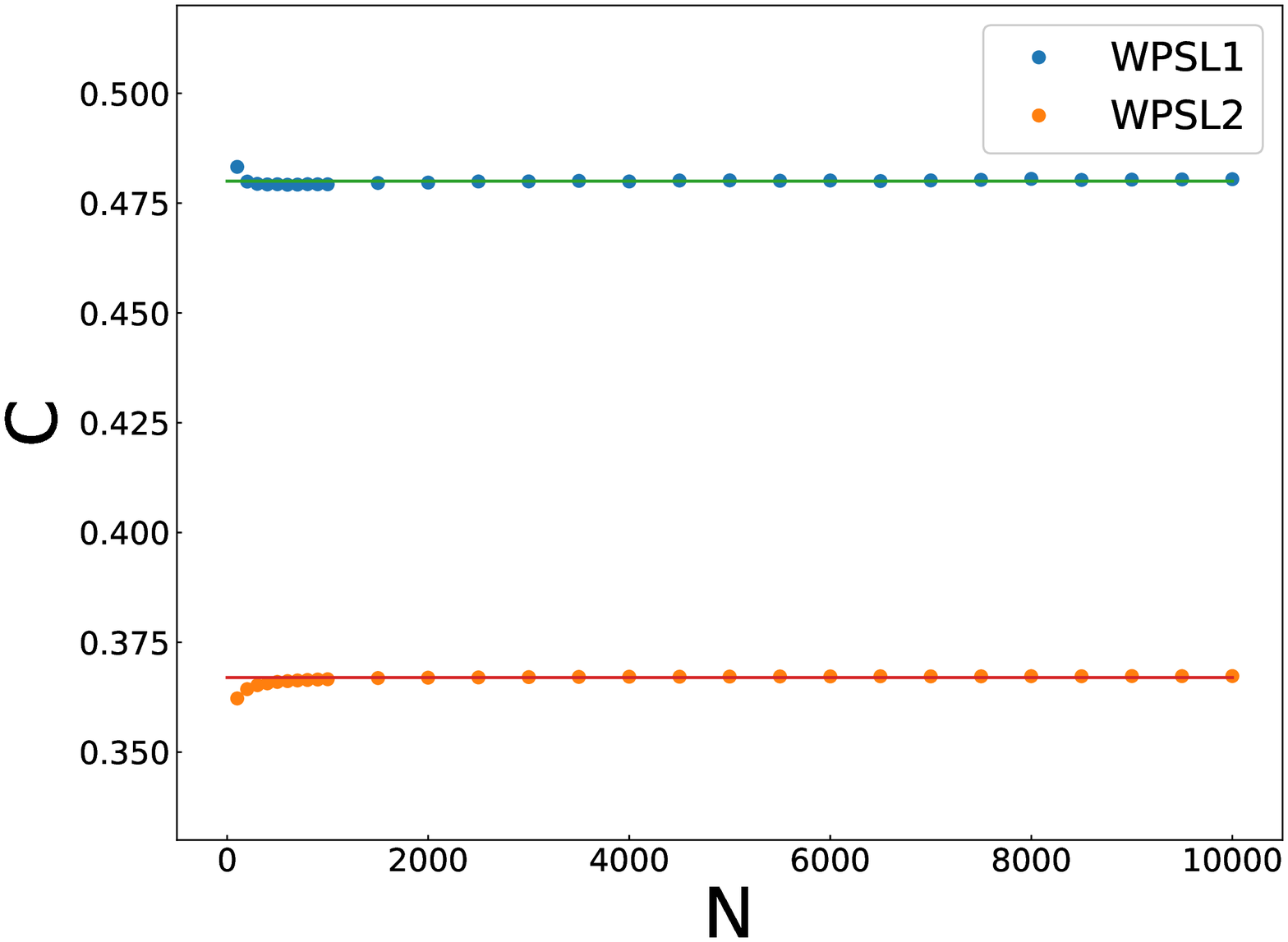}
\label{fig:7}

\caption{
Plot of the clustering coefficient $C$ versus $N$ for WPSL1 and WPSL2 revealing that it is independent
of network size and $C$ is higher for WPSL1 than that for WPSL2.
}

\label{fig:7}
\end{figure}

It has been found that besides the power-law degree distribution, much of the man-made and natural 
networks have two more striking characteristic features 
in common \cite{ref.newman_2}. First, the clustering coefficient $C$, that quantifies
the probability that two nodes having a common neighbor are also neighbors of each other, is high
and independent of system size in the large size limit.
Second, the mean geodesic path length $l$, that describes the shortest mean path between two
arbitrary nodes, is minuscule size as it increases logarithmically with system size. 
These two properties are also known as
the benchmark of the small-world phenomena. The first theoretical
model that can capture the small-world network features was proposed by Watts and Strogatz (WS) in 1998.
However, this model can describe two of the three characteristic features of the real-world networks
with the exception of one, namely the power-law degree distribution.

\begin{figure}
\centering

\subfloat[]
{
\includegraphics[height=4.5 cm, width=4.25 cm, clip=true]
{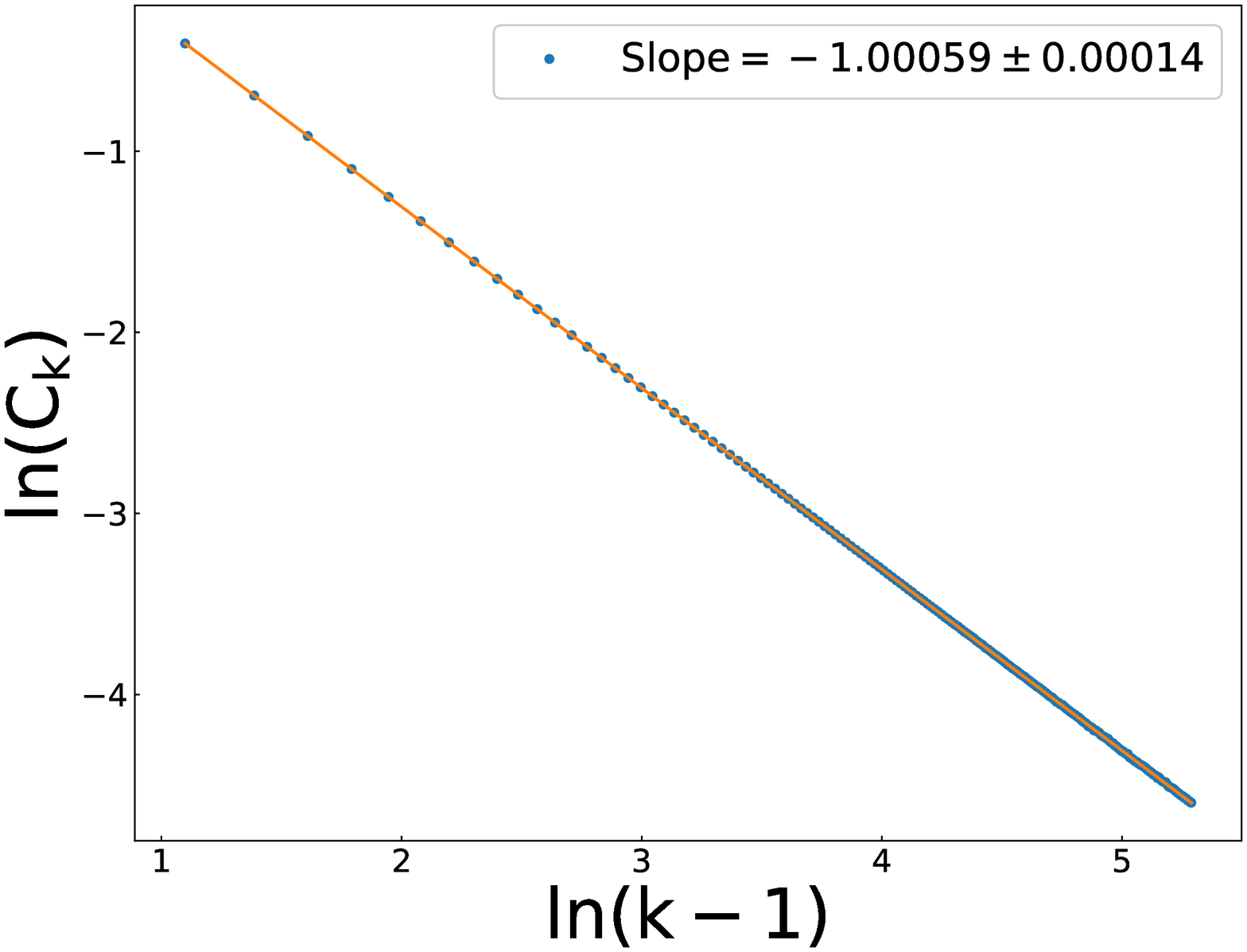}
\label{fig:8a}
}
\subfloat[]
{
\includegraphics[height=4.5 cm, width=4.25 cm, clip=true]
{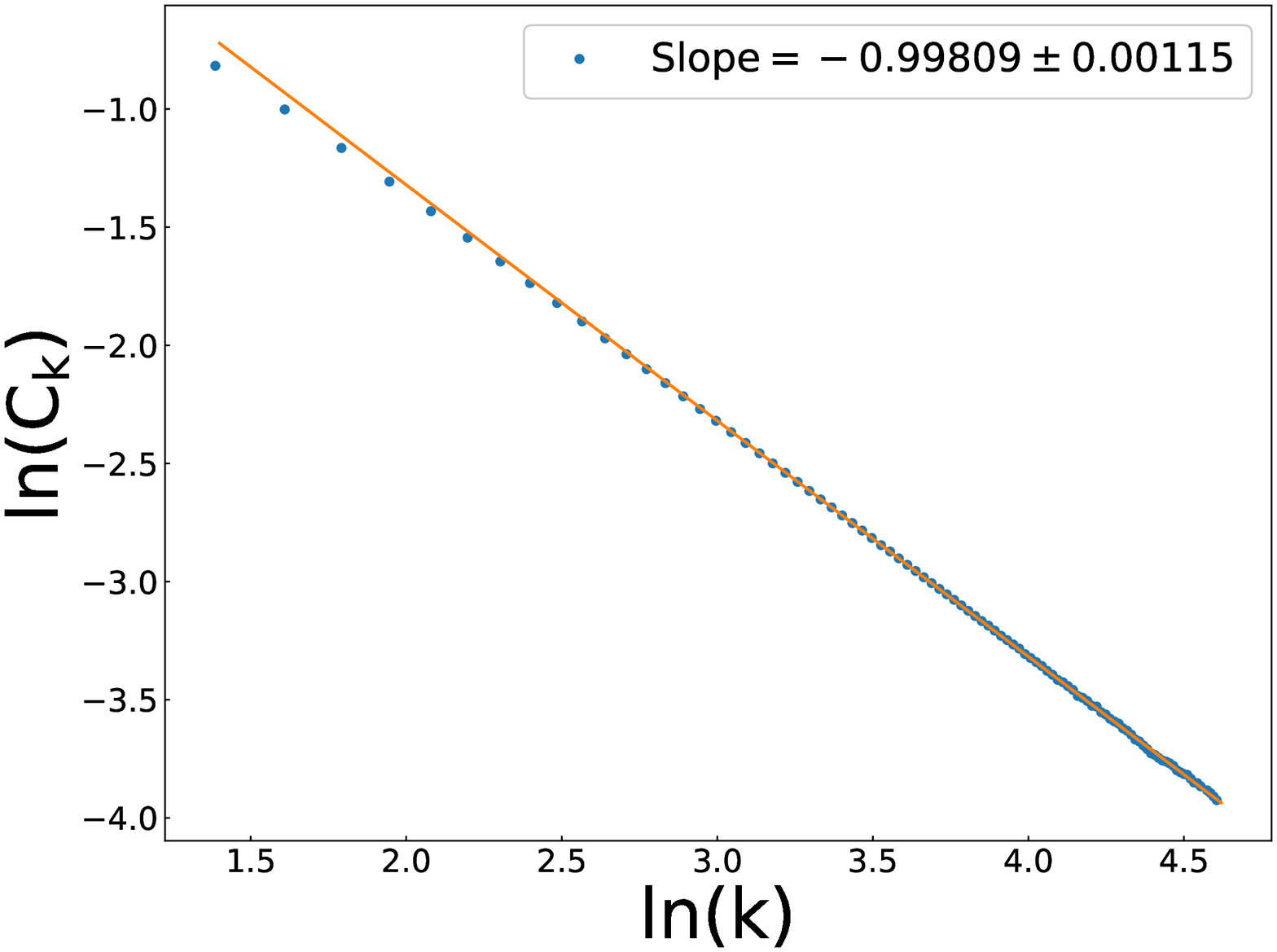}
\label{fig:8b}
}
\caption{Plots of (a) $\ln(C_k)$ versus $\ln(k-1)$  for  WPSL1 and (b) $\ln(C_k)$ versus $\ln(k)$ 
for WPSL2  give straight line 
with slope approximately one in both cases. It suggests clustering coefficient $C_k$
decreases inversely with $k-1$ for WPSL1 and with $k$ for WPSL2 and hence they 
are nested hierarchical networks.} 

\label{fig:8ab}
\end{figure}

We shall now look into the clustering coefficient, also known as transitivity in the social network,  
that describes how likely it is that two friends of a given individual are also friends of each other.
Mathematically, the clustering coefficient $C_i$ of a node $i$ of the network is defined as
the ratio of the number of edges $e_i$ among its $k_i$ neighbors and the number of 
potential edges ${{k_i(k_i-1)}\over{2}}$ among the same neighbors that can exist i.e.,
\begin{equation}
C_i={{e_i}\over{{{k_i(k_i-1)}\over{2}}}}.
\end{equation}
The average of this probability over all the nodes in the network is called the 
clustering coefficient $C$ of the network and hence
\begin{equation}
C={{1}\over{N}}\sum_i^N C_i.
\end{equation}
Except the tree network where $C=0$, the clustering coefficient $C$ is often non-zero and often
it starts high and then decreases like in the  Barab\'{a}si-Albert network. However, sometimes $C$
also has high value and retain the same value independent of the system size. 
The plots of clustering coefficient for WPSL1 and WPSL2 are shown in Fig. (\ref{fig:7}) as a function of lattice size $N$. The mean clustering coefficient for WPSL1 and WPSL2 are  $0.48$ and $0.367$
and they are independent of the system size. 


Yet another quantity of interest
is to find the clustering coefficient $C_k$ of the nodes which have degree $k$ and check if it 
decays following a power-law or not as a function of $k$. Many network models consist of numerous small communities that form larger communities, which may again combine into even larger communities; this
process can continue up to several stages. The quantitative measure
of this nested hierarchical community is provided by finding the dependence of the clustering coefficient
$C_k$ on the node which have degree $k$ \cite{ref.ravasz,ref.hierarchical_ravasz,ref.hierarchical_Dorogovtsev}.
Interestingly, we find nested hierarchical communities as $C_k$ of WPSL1
decreases exactly as 
\begin{equation}
C_k={{2}\over{k-1}},    
\end{equation} 
and hence in the large $k$ limit it decays 
following a power-law $C_k\sim k^{-\nu}$ with $\nu=1$ as shown in Fig. (\ref{fig:8a}). 
The same result, in the large $k$ limit, is also true for WPSL2 as shown in Fig. (\ref{fig:8b}).


\begin{figure}
\centering

\subfloat[]
{
\includegraphics[height=4.5 cm, width=4.25 cm, clip=true]
{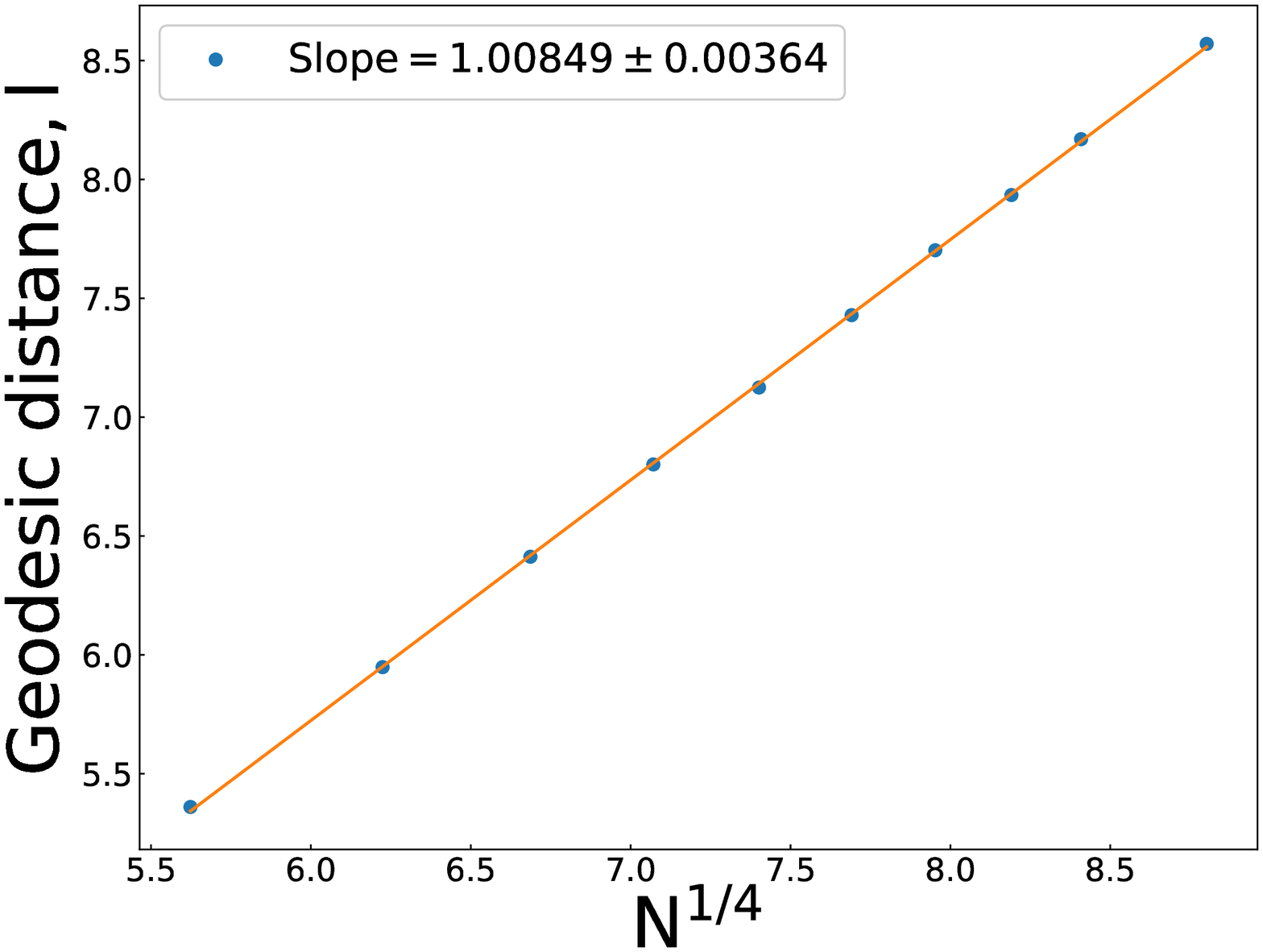}
\label{fig:9a}
}
\subfloat[]
{
\includegraphics[height=4.5 cm, width=4.25 cm, clip=true]
{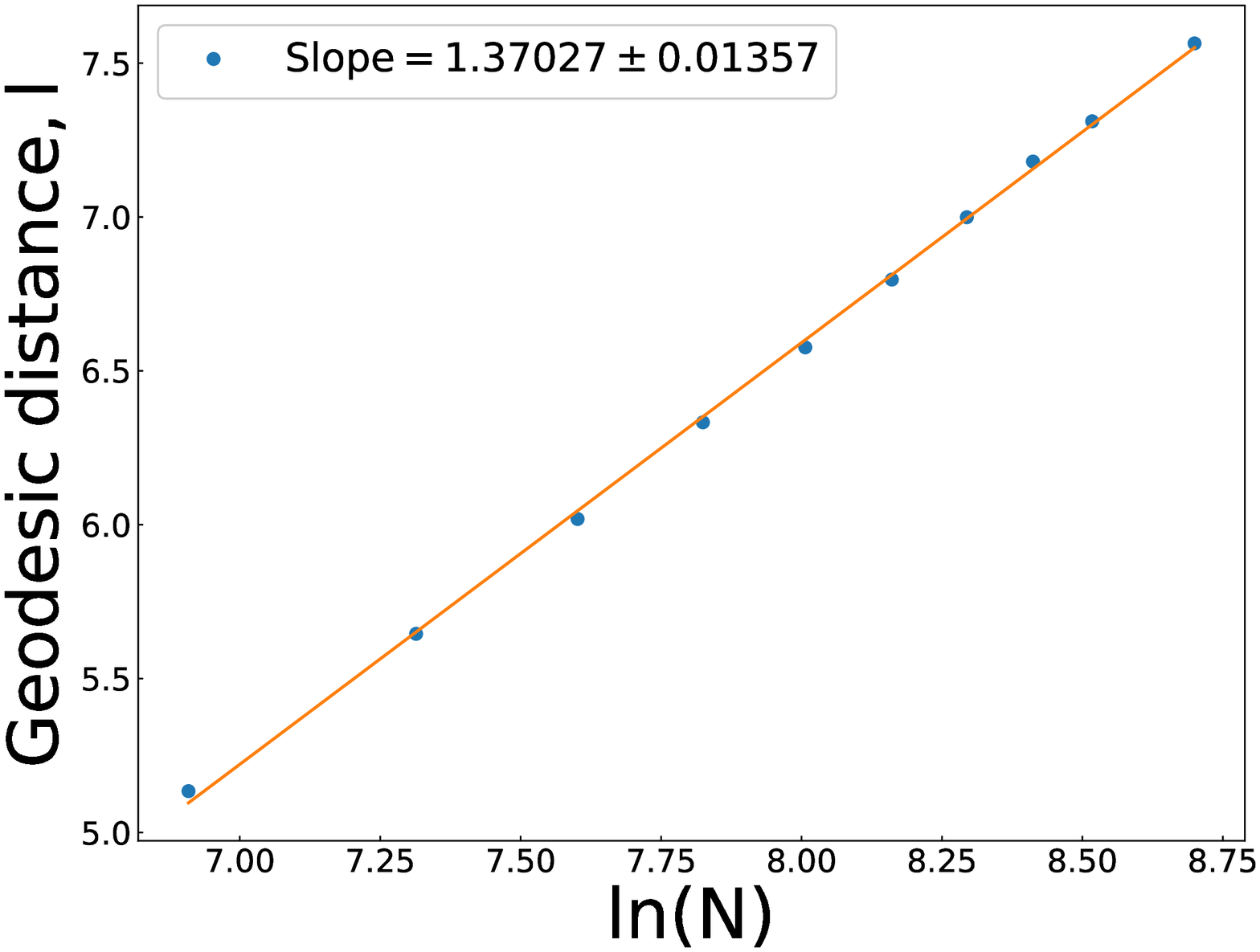}
\label{fig:9b}
}

\caption{Plots of the (a) mean geodesic distance $l$ vs $N^{1/4}$ for WPSL2 and (b) $l$ vs $ln(N)$ are shown and and straight lines are found with slope $1.00$ and $1.37$ respectively.} 

\label{fig:9ab}
\end{figure}

Now we may look at the mean geodesic distance of WPSL1 and WPSL2. 
It is defined as the average of the shortest paths among all the 
connected pairs of nodes in the network i.e.,
\begin{equation}
\label{eq_geodesic_length}
l={{1}\over{2}}{{1}\over{\left( \begin{array}{c}
N \\ 2 \end{array}\right )}}\sum_{i,j}d_{ij}={{1}\over{N(N-1)}}\sum_{i,j}d_{ij},
\end{equation}
where $d_{ij}$ is the shortest path length between node $i$ and $j$ and factor $1/2$ is included
to take care of double counting.  Generally, for regular lattice, the geodesic distance scales as $l\sim N^{1/d}$, where $d$ is the dimension of the support of the lattice. 
Note that the logarithmically slow growth of $l$
is regarded as one of the benchmark of the small-world phenomena as it embodies the 
highly counter-intuitive idea that everyone in the world is on the average only six links of acquaintance away from everyone else. 
The idea of six degrees of separation was first conceived by Hungarian author F. Karinthy and later proved experimentally by S. Milgram \cite{ref.milgram}.

Interestingly, for WPSL2 we find that the mean geodesic distance scales as $l\sim N^{\beta}$ with $\beta=1/4$. To prove this we plot mean geodesic distance $l$ as a function of $N^{1/4}$ and find a straight line with a slope $1.00$ as shown in the Fig.(\ref{fig:9a}). This is reminiscent of the geodesic
path length of $d$ dimensional regular hypercubic lattice where $l$ grows as $N^{1/d}$
\cite{ref.newman}. It suggests that WPSL2 is not really a small-world. However, for WPSL1, 
we find that mean geodesic distance scales as $l\sim ln(N)$. 
We plot the mean geodesic distance as a function of $ln(N)$ which results in a straight line with a slope $1.37$, as shown in Fig.(\ref{fig:9b}). It means that the 
mean geodesic path length grows much slowly for WPSL1 than WPSL2. 
Thus, high clustering co-efficient and logarithmically small geodesic mean path length makes the 
WPSL1 a perfect small-world network.

\section{Summary and Conclusions}

In this article, we have first briefly revisited a class of weighted planar stochastic lattice which we re-named here as WPSL2. This lattice have rich topological and geometric properties since its dual is a scale-free network and geometrically it is a multi-multifractal. However, one of the critique of WPSL2 though is that the exponent $\gamma=5.66$ of the degree distribution is far too high compared to that of the most real-life or man-made networks. Also, the small world properties of WPSL2 have never been investigated.
To this end, we have found that it has small clustering coefficient, compared to WPSL1,
which is independent of system size
but the mean geodesic path length grows following a power-law thus revealing that it is not really
a small-world. We proposed yet another weighted planar stochastic lattice namely WPSL1 in which the generator divides the initiator
randomly either horizontally or vertically into two blocks with equal probability 
instead of four blocks and it is once again applied sequentially exactly in the same way 
as we have done for WPSL2. It has been shown analytically that the dynamics
of the system is governed by infinitely many non-trivial conservation laws and
we have verified it numerically too. We have shown that the ensemble average numerical value of each
conserved quantity, although different for each independent realization, 
is always equal to one. Except the trivial mass conservation, each of the 
non-trivial conserved quantities is distributed in the lattice in such a fashion that the distribution can be 
best understood as multifractal. Since there are infinitely many conserved quantities
for both WPSL1 and WPSL2, there are infinitely many multifractal measures too. 

Interestingly, the dual of the weighted planar stochastic lattice can be mapped
as a network if we replace the center of each cell by node and common boarder between cells
as the link between the two nodes. The corresponding network can also be regarded as
the dual of the WPSL. Furthermore, We have found that like WPSL2, the dual of the WPSL1
too self-organize into a scale-free network and its exponent $\gamma$ is much 
closer to the most real-life and man-made network than that of the WPSL2. 
We have found that the clustering coefficient $C$ for both the cases are quite
high and remain independent of the system size. However, $C$ of the WPSL1 is much higher than that of the WPSL2. Besides, the clustering co-efficient $C_k$ of the nodes which have degree $k$ decreases 
inversely with $k$ in the large $k$ limit in both cases suggesting that they are also hierarchical networks. 
On the other hand, we have found that the mean geodesic path length in WPSL1 is minuscule in
size as it grows logarithmically with system size ($l\sim ln(N)$). However, the mean geodesic path
length for WPLS2 grows algebraically with system size ($l\sim N^{1/4}$) reminiscent of that of
hypercubic lattice. It means that WPSL1 truly is a small world network but not WPSL2.

To summarize, we have proposed a lattice, namely WPSL1, that has many interesting properties. Firstly, it has infinitely many conserved quantities each of which is a multifractal measure. Its dual is a scale-free
network as its degree distribution exhibits a power-law with exponent quite close to natural and man-made
scale-free networks. It is also a small-world as its mean geodesic path length grows logarithmically with
system size and its clustering co-efficient is significantly high ($C=0.48$). Besides, we find
that the clustering coefficient $C_k$ of the nodes which have degree $k$ decays inversely 
with $k-1$ revealing nested hierarchical community. All
these properties make it a lattice that can potentially be used as a skeleton to study
diffusion, percolation etc. One of the central prediction of thermal and probabilistic 
models for continuous phase transition is that their critical behaviors 
depend only on the dimension of the lattice or skeleton and not on the specific choice of lattice. 
For instance, the critical exponents of percolation on all kind of two dimensional lattices are
found to belong to the same universality class. However, recently we have shown this is no longer the case if the lattice, namely WPSL2, has a power-law degree distribution. Now that we know WPSL1 too
is not only scale-free but also hierarchical small-world network with exponent of the degree distribution
much lower than that of WPSL2, it would be interesting to investigate the nature of universality class of percolation on WPSL1. We hope to do this in our future endeavor.

\end{document}